\documentclass[11pt]{article}

\newcommand{\Vector}[1]{\mbox{{\boldmath $#1$}}}
\newcommand{\AlgoName}[1]{\textsc{#1}}

\usepackage{textcomp}
\usepackage[english]{babel}
\usepackage{amsmath}
\usepackage{amsfonts}
\usepackage{enumerate}
\usepackage{framed}
\usepackage{graphicx}
\usepackage{textcomp}
\usepackage{listings}
\usepackage{float}
\usepackage{caption}
\usepackage{subcaption}
\usepackage{hyperref}
\usepackage{rotating}
\usepackage{arydshln}
\begin{document}

\title{A Note on\\
  Experiments and Software \\For Multidimensional Order Statistics}
\author{David L. Woodruff (1) and Stefan Zillmann (2) \\
((1)  University of California Davis (2) University of Duisburg-Essen)
}

\maketitle

\begin{abstract}
In this note we describe experiments on an implementation of two methods
proposed in the literature for computing regions that
correspond to a notion of order statistics for multidimensional data.
Our implementation, which
works for any dimension greater than one, is the only that we know
of to be publicly available. Experiments run using the software
confirm that half-space peeling generally gives better results
than directly peeling convex hulls, but at a computational cost.
\end{abstract}

%\begin{keyword}
%Multivariate order statistics \sep half-space peeling \sep convex hull peeling.
%\end{keyword}

\section{Introduction}

For one dimensional data, the definition of order statistics
renders their computation mostly a matter of sorting.
There is no direct analog for higher dimensions.
In this note we describe experiments on an implementation of two methods
proposed in the literature for multivariate data and a third method
based on multivariate normality that we use as a benchmark.
The software is available for download as open source.

Throughout our discussion, assume that we are presented
with $n$ data points in dimension $p$, which we refer
to as $\Vector{x}_{i} \in \mathbb{R}^{p}, \; i = 1,\dots,n$. For $p=1$,
where there is an obvious order on the data,
it will be useful for
our discussion to use the parameter $\alpha$
that specifies that we want the data point with
the property that $\alpha n$ data points have larger values,
which fits well with our application to
multivariate data.

The methods we consider
share the property that they produce convex {\em regions}
that contain $(1-\alpha)n$ points. Since we do not
want to delve into issues of interpolation, in the sequel
we sometimes use the symbol $\hat{\alpha}$ as the
{\em realized} value of $\alpha$.
Of course, if our interest was in $p=1$ we would have
implemented interpolation, but in higher dimension many
things are not so obvious. 

The methods we employ are a special case of
a more general family related
to notions of data {\em depth} that
have attracted some theoretical attention
(see, e.g., \cite{donoho92,einmahl2015bridging,zuo2000}). However, 
our interest here is strictly computational and
we are not primarily interested in the median or in
extremal values, but rather in a multivariate
version of the concept of order statistics.

Tukey \cite{tukey75} introduced the idea of
depth based on convex hulls demarcated by the intersection
of half-spaces defined by the data. Eddy \cite{eddy82} provided
some probabilistic interpretation and a concise
description of an algorithm, which we employ.
See \cite{WellerEddy} for an updated review of this
method and its application.
To distinguish our implementation of the algorithm,
we use the name \AlgoName{halfspace}. Our implementation, which
works for any dimension $p > 1$ is the only that we know
of to be publicly available.

Eddy used the title ``Convex Hull Peeling,'' but a more
direct use of that name would seem to have been employed
by Barnett \cite{barnett76} where convex hulls defined by the
data are successively peeled to define data depth. This algorithm
was implemented by McDermott and Lin \cite{mcdermott} because
they note (and we confirm computationally) that computation
of half-space intersections is impractical with a serial
implementation applied to massive data sets. We
refer to our implementation of \textit{direct convex hull peeling}
as \AlgoName{direct}.

For benchmarking purposes, we implement a third method that we refer to as \AlgoName{mahal}.
This method is based on a multivariate normal model of the data and defines
the points for parameter value $\alpha$ as the $(1-\alpha)n$ points
whose Mahalanobis distance \cite{mahal} from the mean of the data are lowest.
This method does not share the robustness properties of \AlgoName{halfspace}
(e.g., to the extent that it could provide something like a median, it would
always be very close to the mean), but it provides a useful benchmark, particularly
for simulated data drawn from a multivariate normal population.

A closely related issue that is not studied here, is the issue
of determining the depth of a given point (see, e.g.,
\cite{aloupis2002,Bremner2008,dyckerhoff} . That is, given a point,
find out in which depth region it lies.  This determination does not
require enumerating the regions, but in the case of half-space
depth, does require finding the relevant half-spaces.
An efficient implementation is provided by \cite{Paindaveine2012840},
who describe an implementation of the methods proposed in
\cite{hallin2010}. They stop short of enumerating the points that form
the intersection of half-spaces and then peeling them, but they
provide a very efficient method of finding the halfspaces, which
enables rapid determination of the depth of a point. The R package
{\em depth} \cite{depth} provides numerous methods for finding
depth and enumerates depth regions for $p=2$. 

We proceed as follows. The next section provides a description of
the algorithms that we tested. In Section~\ref{sec:experiments} we
describe some experiments that compare the methods. The final section
offers concluding remarks.

\section{Algorithm Descriptions}

In all three
algorithms, the region is defined as a convex hull.
The first two algorithms make use of iterative peeling where
points on the convex hulls that are identified are removed from the active
data; the third does not. 
Our algorithms
are implemented in Python.
We use the \textit{pyhull} module, which is
a Python wrapper for the qhull (\url{http://www.qhull.org/})
implementation of the Quickhull algorithm \cite{Barber96thequickhull},
to identify points in a convex hull and to produce the intersection of the
half-spaces.

\subsection{Halfspace Depth Convex Hull Peeling Algorithm}
\label{sec:halfspace}

We implemented the halfspace depth convex hull peeling algorithm
-- \AlgoName{halfspace} -- as
described by Eddy \cite{eddy82} with $n$, $p$, and $\alpha$
as inputs, all $n$ points initially classified as {\em active},
and the iteration counter $k$ initially set to zero:

\textbf{Step 0.} Find the hyperplanes defined by all combination of
$p$ active points. We count the number of points in each of the
induced halfspaces (ignoring the points lying directly on the plane.)
We then index the hyperplane by the minimum of these two numbers.

\textbf{Step 1.} Generate the intersection of all
halfspaces with index $k$. In the first iteration with $k=0$, this is the
same as calculating the convex hull $\mathbb{K}_k=conv(N_k \cup v_k)$
of all the available points $N_k \cup v_k$. This is the convex hull of
the whole data-set of points, with $v_0 = \{\}$ and $N_0 = N$. For
subsequent iterations this is the intersection of all halfspaces
defined by the hyperplanes with index $k$.

\textbf{Step 2.} If the number of points contained in intersection of
halfspaces is less than $(1-\alpha)n$, terminate. If not, remove the
points on the convex hull from the active set, increase $k$ by one,
and return to Step 1.

\textbf{Output:} When we reach the
$1 - \alpha$ quantile, we compare the last two hulls with each other
and output the hull with $\hat{\alpha}$ closest to $\alpha$

\begin{figure}[H]
	\centering
	\begin{minipage}{.5\textwidth}
		\centering
		\includegraphics[width=\textwidth]{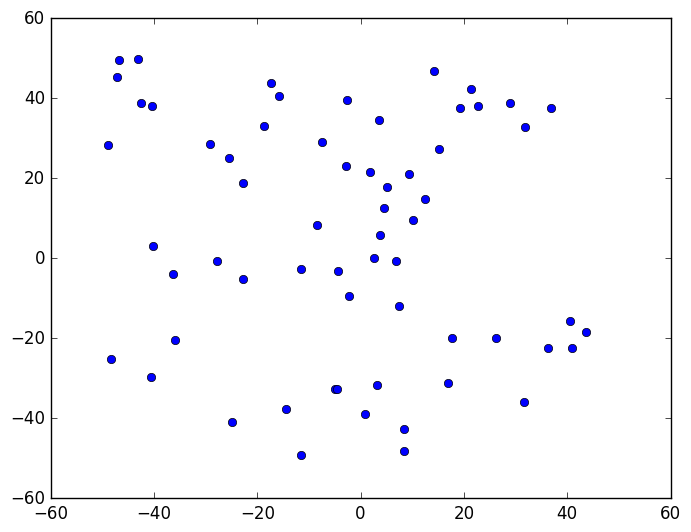}
		%\captionof{figure}{The original data set}
		%\label{}
	\end{minipage}%
	\begin{minipage}{.5\textwidth}
		\centering
		\includegraphics[width=\textwidth]{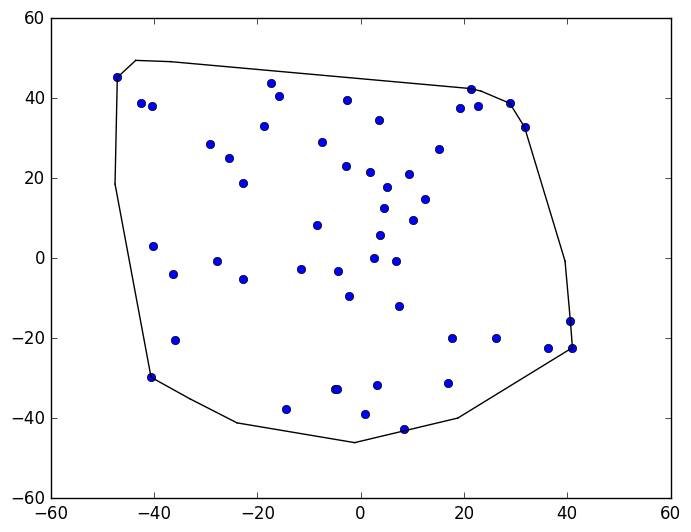}
		%\captionof{figure}{convex hull peeling result ($\alpha=0.1$)}
		%\label{plot2}
	\end{minipage}
	\caption{The result of halfspace depth peeling ($\alpha=0.1$) on an example data set.}
\end{figure}

\subsection{Direct Convex Hull Peeling Algorithm}
\label{sec:direct}

This algorithm -- \AlgoName{direct} -- iteratively calculates the
convex hull of the active set of points and then removes all the
points on the outside faces of the hull. When there are $(1 - \alpha)n$
or fewer points in the active data set, the algorithm stops and
outputs the points in the hull that contains the number of points closest to $(1-\alpha)n$.

\begin{figure}[H]
	\centering
	\begin{minipage}{.5\textwidth}
		\centering
		\includegraphics[width=\textwidth]{all.png}
		%\captionof{figure}{The original data set}
		%\label{}
	\end{minipage}%
	\begin{minipage}{.5\textwidth}
		\centering
		\includegraphics[width=\textwidth]{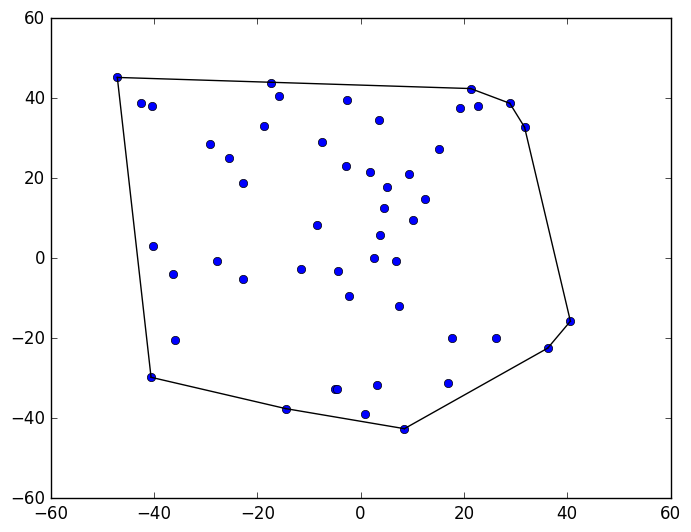}
		%\captionof{figure}{convex hull peeling result ($\alpha=0.1$)}
		%\label{plot2}
	\end{minipage}
	\caption{The result of \textit{direct convex hull peeling} ($\alpha=0.1$) on an example data set.}
\end{figure}

\subsection{Mahalanobis Distance Algorithm}
\label{sec:mahal}

The \textit{Mahalanobis distance} algorithm -- \AlgoName{mahal} --
first uses the mean, $\overline{\Vector{x}}$, and the covariance matrix,
$\hat{\Sigma}$ of the full data set to
  calculate the Mahalanobis distance of each point from the mean.
Assuming the data are given as column vectors, 
a data point $\Vector{x}$ has distance
$$
(\Vector{x}-\overline{\Vector{x}})^{T}
\Sigma^{-1}
(\Vector{x}-\overline{\Vector{x}}).
$$
The convex
hull of the innermost $\lfloor (1 - \alpha)n \rceil$ points provides
the desired region, where $\lfloor ... \rceil$ denotes rounding to the
nearest integer.

\begin{figure}[H]
	\centering
	\begin{minipage}{.5\textwidth}
		\centering
		\includegraphics[width=\textwidth]{all.png}
		%\captionof{figure}{A prediction interval plot}
		%\label{plot}
	\end{minipage}%
	\begin{minipage}{.5\textwidth}
		\centering
		\includegraphics[width=\textwidth]{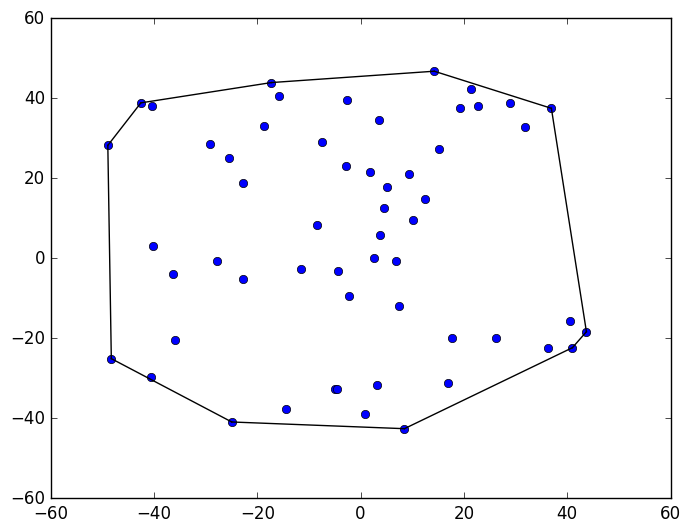}
		%\captionof{figure}{mahalanobis distance algorithm ($\alpha=0.1$)}
		%\label{plot2}
	\end{minipage}
	\caption{The result of the \textit{mahalanobis distance algorithm} ($\alpha=0.1$) on an example data set.}
\end{figure}

\section{Computational Experiments}
\label{sec:experiments}

We conduct experiments using $\alpha=0.1, 0.5, 0.9$.  Bear in mind
that while for $p=1$, the value of $\alpha=0.5$ would be the median,
the corresponding concept of centrality for $p>1$ is the ``inner-most
peel'' which will usually correspond to $\hat{\alpha} < 0.1$ for all
but small values of $n$; however, our interest is in order statistics
not centrality. To measure the quality of shapes that are intended to
correspond to order statistics, we consider both the {\em skill} and
{\em sharpness}. For a fixed value of $\alpha$, the sharpness is
simply the area (for $p=2$) or volume. We measure the skill using the
{\em error}. For each replicate we find the hull with a value of
$\hat{\alpha}$ close to $\alpha$ for the {\em training} data, then we
examine test data sets (100 sets in the simulations) and for each test
data set we use the difference between the fraction of points inside
the hull and $(1-\hat{\alpha})$ as the error.  For the practitioners
that we work with, skill seems to be the more important of the two
measures, but for roughly comparable skill, one would obviously prefer
improved sharpness.

\subsection{Simulated Data}
\label{sim}

The simulated data sets we used are generated from multivariate normal
distributions with mean zero. For each value of $p$, we generate data
sets using one of two covariance matrices and also data sets that have
a mixture of data from these two distributions. We use 10 replicates
for each value of $n$, $p$, and covariance structure. A replicate
consists of a training data set and 100 data sets used to test the
skill of the regions found by each method.

We use data two covariance matrices that we refer to as $A$ and $B$ to generate
data. A third type of data has $\frac{n}{2}$ points distributed
N($\Vector{0},A$) and $\frac{n}{2}$ points distributed N($\Vector{0},
B$).

\subsubsection{Two-dimensional data}

For $p=2$, we use these covariance matrices to simulate data:

\begin{center}
$A_{2 \times 2} = \begin{bmatrix} 1 & 0.6 \\ 0.6 & 1 \end{bmatrix}$. And $B_{2 \times 2} = \begin{bmatrix} 5 & -2 \\ -2 & 5 \end{bmatrix}$
\end{center}

In \autoref{big_table}, we use $\mu$ to refer to the mean error of all 10 replicates and 100 tests for a particular covariance
structure. The standard
deviation $\sigma$ and the minimum ($min$) and maximum ($max$) of this error
over the replicates is also displayed.
Note that the {\sc direct} algorithm has the
same values in the experiments that only change the covariance
matrix. This is due to the fact that all experiments have the same
uniform random numbers as input to make them comparable.

For detailed results, refer to \autoref{detail_1a}, \autoref{detail_1b}, \autoref{detail_2a} and \autoref{detail_2b}.
Overall \AlgoName{halfspace} has the best skill, which it seems to achieve
by producing, on average, slightly larger areas (volumes).

\subsubsection{Three-dimensional data}

For $p=3$, we use these covariance matrices:
\begin{center}
	$A_{3 \times 3} = \begin{bmatrix} 1 & 0.6 & 0.6 \\ 0.6 & 1 & 0.6 \\ 0.6 & 0.6 & 1 \end{bmatrix}$ and $B_{3 \times 3} = \begin{bmatrix} 5 & -2 & -2 \\ -2 & 5 & -2 \\ -2 & -2 & 5 \end{bmatrix}$.
\end{center} 
A summary of results can be seen in \autoref{3d_table}.

The realized
$\hat{\alpha}$ for $n=200$ and $p=3$ is much further away from the desired
alpha value than any experiment for $p=2$. The realized $\hat{\alpha}$ for
$COV=A_{3 \times 3}$ with $\alpha = 0.1$ for example varies between
$0.0$ and $0.2$ for the \textit{halfspace depth convex hull peeling}
algorithm, which is a difference of $0.1$. Even for the $n=100$ case
with $COV=A_{2 \times 2}$ and $\alpha = 0.1$, the \textit{halfspace
  depth convex hull peeling} algorithm in the two dimensional case
differs only by $0.015$ at most, which confirms that increasing $p$
results in a need for more data points. These results also confirm
that the \textit{halfspace depth convex hull peeling} tends to
produce the better skill results.

\subsubsection{Runtime} % Dave, Should i rename \textit{mahalanobis distance} everywhere in the paper to \AlgoName{mahal} etc? So that we keep one style? DLW: if we are being really careful, then \textit{mahalanobis distance} should be the method and \AlgoName{mahal} should be our implementation of it.

One major drawback of the \textit{halfspace depth convex hull peeling}
algorithm is runtime.  \autoref{time_table} shows how the different
algorithms compare using our serial implementation.

While the
overall construction time for the \textit{mahalanobis distance} and
\textit{direct convex hull peeling} algorithms is about the same
for $10$ replicates of any covariance matrix with $n=100$, the \textit{halfspace depth convex hull peeling}
algorithm needed about $3$ times as many seconds in all of the
equivalent experiments. 
Also, the time required for \textit{halfspace depth convex hull peeling}
increases drastically. For $n=500$, the \textit{direct convex hull
  peeling} and \textit{mahalanobis distance} algorithms stay under
$20$ seconds run time for all trials. The \textit{halfspace depth
  convex hull peeling} meanwhile reaches more than $3000$ seconds
($50$ minutes) in this case. This problem only increases with higher
dimensional data.
As Eddy
notes, this requires O($n^{p+1}$) time, which we see can be
significant for our implementation.

The most time consuming part is the creation of all
halfspaces and checking how many points are on either side.  It would be trivial to parallelize this; however, it would be not so trivial to parallelize
finding the intersection, which would eventually become the
computational bottleneck.

\section{Conclusions}

We have described experiments on an implementation of methods proposed
in the literature for computing regions that correspond to a notion of
order statistics for multidimensional data.  The software is publicly
available on github under the name morderstats to support use in
practice or additional experiments.

Experiments that we report on here
confirm that half-space peeling generally gives better results
than directly peeling convex hulls, but at a computational cost.
%%% DLW:
%%%Parallel algorithm: index the points and
%%%then assign points to processors; a processor
%%%should form all planes for the points below in the list.
%%%Using this allocation scheme, processor loads can be balanced pretty easily.
%%%%
One potential avenue for further research is the use of faster methods
that could be used to find the planes (see, e.g.,
\cite{Paindaveine2012840}) or one could parallelize the method we use.
Then computing the intersections would become the bottleneck and a potential topic
for further research.

Our work was motivated by wind power prediction intervals
\cite{pes2017} where order statistics offer the advantage that they
are easy to explain to all stakeholders. They are also useful as a
benchmark against which to compare model-based methods. We are now
working to simultaneously consider adjacent hours and multiple
sources of renewable energy. These
applications highlight the need for ongoing research in computational
methods for multidimensional order statistics.

\section*{Acknowledgments}

Helpful comments were provided by David M.\ Rocke. This work was
funded in part by the Bonneville Power Administration.

\section*{References}
\bibliographystyle{plain}
\bibliography{hullpaper}

\begin{thebibliography}{10}

\bibitem{aloupis2002}
Greg Aloupis, Carmen Cort{\'e}s, Francisco G\'{o}mez, Michael Soss, and
  Godfried Toussaint.
\newblock Lower bounds for computing statistical depth.
\newblock {\em Comput. Stat. Data Anal.}, 40(2):223--229, August 2002.

\bibitem{Barber96thequickhull}
C.~Bradford Barber, David~P. Dobkin, and Hannu Huhdanpaa.
\newblock The quickhull algorithm for convex hulls.
\newblock {\em ACM Transactions on Mathematical Software}, 22(4):469--483,
  1996.

\bibitem{barnett76}
V.~Barnett.
\newblock The ordering of multivariate data.
\newblock {\em Journal of the Royal Statistical Society, Series A},
  139:318--–352, 1976.

\bibitem{Bremner2008}
David Bremner, Dan Chen, John Iacono, Stefan Langerman, and Pat Morin.
\newblock Output-sensitive algorithms for tukey depth and related problems.
\newblock {\em Statistics and Computing}, 18(3):259, 2008.

\bibitem{donoho92}
D.~Donoho and M.~Gasko.
\newblock Breakdown properties of location estimates based half-space depth and
  projected outlyingness.
\newblock {\em Ann. Statist.}, 20:1803–--1827, 1992.

\bibitem{dyckerhoff}
R.~Dyckerhoff and P.~Mozharovskyi.
\newblock Exact computation of the halfspace depth.
\newblock {\em Comput. Statist. Data Anal.}, 98:19--–30, 2016.

\bibitem{eddy82}
W.F. Eddy.
\newblock Convex hull peeling.
\newblock In {\em COMPSTAT 1982 5th Symposium held at Toulouse}, pages 42--47,
  1982.

\bibitem{einmahl2015bridging}
John~HJ Einmahl, Jun Li, Regina~Y Liu, et~al.
\newblock Bridging centrality and extremity: Refining empirical data depth
  using extreme value statistics.
\newblock {\em The Annals of Statistics}, 43(6):2738--2765, 2015.

\bibitem{depth}
Maxime Genest, Jean-Claude Masse, and Jean-Francois Plante.
\newblock depth: Nonparametric depth functions for multivariate analysis.
\newblock \url{https://CRAN.R-project.org/package=depth}, 2017.

\bibitem{hallin2010}
Marc Hallin, Davy Paindaveine, and Miroslav {\v S}iman.
\newblock Multivariate quantiles and multiple-output regression quantiles: From
  $l_1$ optimization to halfspace depth.
\newblock {\em Ann. Statist.}, 38(2):635--669, 04 2010.

\bibitem{mahal}
P.C. Mahalanobis.
\newblock On the generalised distance in statistics.
\newblock In {\em Proceedings of the National Institute of Sciences of India},
  volume~2, pages 49--55, 1936.

\bibitem{mcdermott}
James~P. McDermott and Dennis K.~J. Lin.
\newblock Quantile contours and multivariate density estimation for massive
  datasets via sequential convex hull peeling.
\newblock {\em IIE Transactions}, 39(6):581--591, 2007.

\bibitem{pes2017}
Sabrina Nitsche, C\'{e}sar~A.\ Silva-Monroy, Andrea Staid Jean-Paul Watson,
  Scott Winner, and David~L. Woodruff.
\newblock Improving wind power prediction intervals using vendor-supplied
  probabilistic forecast information.
\newblock In {\em IEEE Power and Energy Society General Meeting (PES)}, pages
  1--5. IEEE, 2017.

\bibitem{Paindaveine2012840}
Davy Paindaveine and Miroslav {\v S}iman.
\newblock Computing multiple-output regression quantile regions.
\newblock {\em Computational Statistics and Data Analysis}, 56(4):840 -- 853,
  2012.

\bibitem{tukey75}
J.~Tukey.
\newblock Mathematics and the picturing of data.
\newblock In {\em Proc. 1975 Inter. Cong. Math., Vancouver, Canada}, page
  523–531. Canad. Math. Congress., 1975.

\bibitem{WellerEddy}
G.B. Weller and W.F. Eddy.
\newblock Multivariate order statistics: Theory and application.
\newblock {\em Annual Review of Statistics and Its Application}, 2(1):237--257,
  2015.

\bibitem{zuo2000}
Yijun Zuo and Robert Serfling.
\newblock General notions of statistical depth function.
\newblock {\em Ann. Statist.}, 28(2):461--482, 04 2000.

\end{thebibliography}

\newpage

\appendix

\section{Tables}

\begin{sidewaystable} [!htbp]
	\begin{center}
	\begin{tabular}{| c | l | l c r | l c r | l c r |}
		\hline
		$\overset{\sigma}{\mu}$ &&& Mahal &&& Direct &&& Halfspace &  \\
		{[min, max]} & $\alpha=$& 0.9 & 0.5 & 0.1 & 0.9 & 0.5 & 0.1 & 0.9 & 0.5 & 0.1 \\
		\hline
		&&&&&&&&&& \\
		& $n = 100$ & $\overset{.0272}{\underset{[.000, .100]}{.0557}}$ & $\overset{.0668}{\underset{[.000, .290]}{.1084}}$ & $\overset{.0520}{\underset{[.000, .280]}{.0939}}$ & $\overset{.0252}{\underset{[.000, .130]}{.0589}}$ & $\overset{.0645}{\underset{[.000, .300]}{.1010}}$ & $\overset{.0506}{\underset{[.000, .250]}{.0918}}$ & $\overset{.0231}{\underset{[.000, .120]}{.0286}}$ & $\overset{.0504}{\underset{[.000, .230]}{.0650}}$ & $\overset{.0461}{\underset{[.000, .230]}{.0566}}$ \\
		\cdashline{3-11} &&&&&&&&&& \\
		$COV = A$ & $n=200$	& $\overset{.0202}{\underset{[.000, .095]}{.0406}}$ & $\overset{.0386}{\underset{[.000, .185]}{.0582}}$ & $\overset{.0337}{\underset{[.000, .175]}{.0609}}$ & $\overset{.0201}{\underset{[.000, .090]}{.0357}}$ & $\overset{.0410}{\underset{[.000, .195]}{.0660}}$ & $\overset{.0311}{\underset{[.000, .165]}{.0637}}$ & $\overset{.0202}{\underset{[.000, .155]}{.0214}}$ & $\overset{.0287}{\underset{[.000, .145]}{.0277}}$ & $\overset{.0245}{\underset{[.000, .125]}{.0334}}$ \\
		\cdashline{3-11}
		&&&&&&&&&& \\
		& $n = 500$ & $\overset{.0144}{\underset{[.000, .070]}{.0248}}$ & $\overset{.0225}{\underset{[.000, .114]}{.0296}}$ & $\overset{.0160}{\underset{[.000, .082]}{.0259}}$ & $\overset{.0142}{\underset{[.000, .066]}{.0214}}$ & $\overset{.0242}{\underset{[.000, .126]}{.0387}}$ & $\overset{.0182}{\underset{[.000, .098]}{.0322}}$ & $\overset{.0115}{\underset{[.000, .054]}{.0119}}$ & $\overset{.0183}{\underset{[.000, .092]}{.0207}}$ & $\overset{.0123}{\underset{[.000, .068]}{.0123}}$ \\
		\hline
		&&&&&&&&&& \\
		& $n = 100$ & $\overset{.0265}{\underset{[.000, .100]}{.0563}}$ & $\overset{.0716}{\underset{[.000, .320]}{.1200}}$ & $\overset{.0552}{\underset{[.000, .280]}{.0989}}$ & $\overset{.0252}{\underset{[.000, .130]}{.0589}}$ & $\overset{.0645}{\underset{[.000, .300]}{.1010}}$ & $\overset{.0506}{\underset{[.000, .250]}{.0918}}$ & $\overset{.0239}{\underset{[.000, .120]}{.0300}}$ & $\overset{.0512}{\underset{[.000, .230]}{.0650}}$ & $\overset{.0432}{\underset{[.000, .210]}{.0557}}$ \\
		\cdashline{3-11}
		&&&&&&&&&& \\
		$COV = B$ & $n=200$	& $\overset{.0198}{\underset{[.000, .095]}{.0406}}$ & $\overset{.0388}{\underset{[.000, .205]}{.0716}}$ & $\overset{.0353}{\underset{[.000, .190]}{.0680}}$ & $\overset{.0201}{\underset{[.000, .090]}{.0357}}$ & $\overset{.0410}{\underset{[.000, .195]}{.0660}}$ & $\overset{.0311}{\underset{[.000, .165]}{.0637}}$ & $\overset{.0200}{\underset{[.000, .155]}{.0208}}$ & $\overset{.0295}{\underset{[.000, .145]}{.0306}}$ & $\overset{.0241}{\underset{[.000, .125]}{.0315}}$ \\
		\cdashline{3-11}
		&&&&&&&&&& \\
		& $n = 500$ & $\overset{.0142}{\underset{[.000, .070]}{.0252}}$ & $\overset{.0236}{\underset{[.000, .120]}{.0370}}$ & $\overset{.0174}{\underset{[.000, .098]}{.0298}}$ & $\overset{.0142}{\underset{[.000, .066]}{.0214}}$ & $\overset{.0242}{\underset{[.000, .126]}{.0387}}$ & $\overset{.0182}{\underset{[.000, .098]}{.0322}}$ & $\overset{.0115}{\underset{[.000, .054]}{.0119}}$ & $\overset{.0180}{\underset{[.000, .092]}{.0197}}$ & $\overset{.0133}{\underset{[.000, .072]}{.0136}}$ \\
		\hline
		&&&&&&&&&& \\
		$COV = A$ & $n=100$ & $\overset{.0244}{\underset{[.000, .100]}{.0575}}$ & $\overset{.0583}{\underset{[.000, .260]}{.1006}}$ & $\overset{.0531}{\underset{[.000, .310]}{.0975}}$ & $\overset{.0262}{\underset{[.000, .120]}{.0581}}$ & $\overset{.0608}{\underset{[.000, .270]}{.0939}}$ & $\overset{.0550}{\underset{[.000, .280]}{.1063}}$ & $\overset{.0233}{\underset{[.000, .130]}{.0280}}$ & $\overset{.0461}{\underset{[.000, .220]}{.0684}}$ & $\overset{.0479}{\underset{[.000, .250]}{.0655}}$ \\
		\cdashline{3-11}
		&&&&&&&&&& \\
		and	& $n=200$        & $\overset{.0200}{\underset{[.000, .090]}{.0397}}$ & $\overset{.0387}{\underset{[.000, .205]}{.0679}}$ & $\overset{.0285}{\underset{[.000, .140]}{.0538}}$ & $\overset{.0227}{\underset{[.000, .100]}{.0420}}$ & $\overset{.0364}{\underset{[.000, .180]}{.0582}}$ & $\overset{.0269}{\underset{[.000, .140]}{.0588}}$ & $\overset{.0184}{\underset{[.000, .100]}{.0251}}$ & $\overset{.0243}{\underset{[.000, .135]}{.0231}}$ & $\overset{.0212}{\underset{[.000, .110]}{.0310}}$ \\
		\cdashline{3-11}
		&&&&&&&&&& \\
		$COV = B$ & $n=500$	& $\overset{.0142}{\underset{[.000, .066]}{.0191}}$ & $\overset{.0246}{\underset{[.000, .120]}{.0400}}$ & $\overset{.0175}{\underset{[.000, .084]}{.0288}}$ & $\overset{.0132}{\underset{[.000, .062]}{.0198}}$ & $\overset{.0234}{\underset{[.000, .118]}{.0398}}$ & $\overset{.0175}{\underset{[.000, .082]}{.0316}}$ & $\overset{.0125}{\underset{[.000, .070]}{.0130}}$ & $\overset{.0167}{\underset{[.000, .088]}{.0196}}$ & $\overset{.0121}{\underset{[.000, .058]}{.0114}}$ \\
		\hline	
	\end{tabular}
        
	\caption{Results of experiments with 10 replicates and 100 test data sets for each in dimension $p=2$. The table shows the mean $\mu$ and standard deviation $\sigma$ for the error for all replicates.  Each replicate involves generation of a data set of $n$ points and calculation the hull according to $\alpha$ for each algorithm. Each iteration within each replicate then generates $n$ new points with the same covariance structure as the original data set had and counts how many points are inside the generated hull. The mean of the difference between all these numbers and each realized alpha, $\hat{alpha}$, then becomes $\mu$, which is the value in the middle of each cell. $\sigma$ is the standard deviation of all these errors and displayed above the $\mu$ value. The minimum error and maximum error are displayed in the form of $[min, max]$ below the $\mu$ value.}
	\label{big_table}

        \end{center}
\end{sidewaystable}

\newpage

\begin{sidewaystable} [!htbp]
	\begin{center}
	\begin{tabular}{| c | l | l c r | l c r | l c r |}
		\hline
		$\overset{\sigma}{\mu}$ &&& Mahal &&& Direct &&& Halfspace &  \\
		{[min, max]} & $\alpha=$& 0.9 & 0.5 & 0.1 & 0.9 & 0.5 & 0.1 & 0.9 & 0.5 & 0.1 \\
		\hline
		&&&&&&&&&& \\
		$COV = A$ & $n = 200$ & $\overset{.0133}{\underset{[.020, .100]}{.0694}}$ & $\overset{.0474}{\underset{[.000, .260]}{.1324}}$ & $\overset{.0414}{\underset{[.050, .270]}{.1602}}$ & $\overset{.0207}{\underset{[.010, .125]}{.0734}}$ & $\overset{.0421}{\underset{[.035, .310]}{.1749}}$ & $\overset{.0406}{\underset{[.065, .325]}{.1801}}$ & $\overset{.0282}{\underset{[.000, .150]}{.0368}}$ & $\overset{.0581}{\underset{[.000, .250]}{.0942}}$ & $\overset{.0505}{\underset{[.000, .240]}{.1261}}$ \\
		\hline
		&&&&&&&&&& \\
		$COV = B$ & $n = 200$ & $\overset{.0153}{\underset{[.020, .100]}{.0695}}$ & $\overset{.0463}{\underset{[.000, .280]}{.1458}}$ & $\overset{.0398}{\underset{[.040, .285]}{.1669}}$ & $\overset{.0207}{\underset{[.010, .125]}{.0734}}$ & $\overset{.0421}{\underset{[.035, .310]}{.1749}}$ & $\overset{.0406}{\underset{[.065, .325]}{.1801}}$ & $\overset{.0208}{\underset{[.000, .125]}{.0206}}$ & $\overset{.0414}{\underset{[.000, .200]}{.0505}}$ & $\overset{.0574}{\underset{[.000, .240]}{.0980}}$ \\
		\hline
		&&&&&&&&&& \\
		$COX = A$ and $COV = B$	        & $n = 200$ & $\overset{.0162}{\underset{[.000, .100]}{.0669}}$ & $\overset{.0365}{\underset{[.020, .245]}{.1361}}$ & $\overset{.0337}{\underset{[.030, .250]}{.1351}}$ & $\overset{.0177}{\underset{[.005, .115]}{.0651}}$ & $\overset{.0369}{\underset{[.020, .230]}{.1251}}$ & $\overset{.0360}{\underset{[.040, .255]}{.1321}}$ & $\overset{.0198}{\underset{[.000, .105]}{.0211}}$ & $\overset{.1255}{\underset{[.000, .505]}{.0848}}$ & $\overset{.0374}{\underset{[.000, .160]}{.0462}}$ \\
		\hline	
	\end{tabular}
	\caption{Results of experiments with $10$ replicates in dimension $p=3$. The table shows the mean $\mu$ and standard deviation $\sigma$ for the error for all replicates.  Each replicate involves generation of a data set of $n$ points and calculation of the hull according to $\alpha$ for each algorithm. Each of the 100 test data sets within each replicate then generates $n$ new points with the same covariance matrix the original data set had and counts how many points are inside the generated hull. The mean of the difference between all these numbers and each realized alpha then becomes $\mu$, which is the value in the middle of each cell. $\sigma$ is the standard deviation of all these errors and displayed above the $\mu$ value. The minimum error and maximum error are displayed in the form of $[min, max]$ below the $\mu$ value.}
	\label{3d_table}
\end{center}
\end{sidewaystable}

\newpage

\begin{table} [!htbp]
	\begin{center}
		\hspace*{-2cm}  
		\begin{tabular}{| c c | r r r | r r r | r r r |}
			\hline
			& $n = 100$ && Mahal &&& Direct &&& Halfspace &  \\
			& $p=2$ & 0.9 & 0.5 & 0.1 & 0.9 & 0.5 & 0.1 & 0.9 & 0.5 & 0.1 \\
			\hline
			 & realized alpha ($\mu$)& 0.900 & 0.500 & 0.100 & 0.899 & 0.509 & 0.083 & 0.904 & 0.504 & 0.090 \\
			& realized alpha ($\sigma$)& 0.000 & 0.000 & 0.000 & 0.0166 & 0.034 & 0.013 & 0.007 & 0.008 & 0.016 \\
			 		& too many points ($\mu$)& 0.080 & 0.077 & 0.019 & 0.029 & 0.084 & 0.026 & 0.369 & 0.319 & 0.122 \\
			$COV = A$   & too many points ($\sigma$)& 0.133 & 0.129 & 0.035 & 0.045 & 0.156 & 0.048 & 0.339 & 0.357 & 0.148 \\
			 & too few points ($\mu$)& 0.920 & 0.905 & 0.967 & 0.970 & 0.897 & 0.957 & 0.631 & 0.640 & 0.819 \\
			& too few points ($\sigma$)& 0.133 & 0.148 & 0.053 & 0.044 & 0.172 & 0.085 & 0.339 & 0.365 & 0.208 \\
			 & volume ($\mu$)& 0.245 & 3.097 & 9.150 & 0.225 & 2.571 & 9.360 & 0.447 & 3.264 & 10.213 \\
			& volume ($\sigma$)& 0.141 & 0.647 & 1.073 & 0.091 & 0.555 & 1.344 & 0.187 & 0.779 & 1.526 \\
			\hline
			& realized alpha ($\mu$)& 0.900 & 0.500 & 0.100 & 0.899 & 0.509 & 0.083 & 0.902 & 0.506 & 0.091 \\
			& realized alpha ($\sigma$)& 0.000 & 0.000 & 0.000 & 0.017 & 0.034 & 0.013 & 0.004 & 0.010 & 0.017  \\
				        & too many points ($\mu$)& 0.080 & 0.062 & 0.017 & 0.029 & 0.084 & 0.026 & 0.369 & 0.330 & 0.092 \\
			$COV = B$ & too many points ($\sigma$)& 0.166 & 0.129 & 0.030 & 0.045 & 0.156 & 0.048 & 0.340 & 0.356 & 0.096 \\
			& too few points ($\mu$)& 0.920 & 0.910 & 0.970 & 0.970 & 0.897 & 0.957 & 0.631 & 0.632 & 0.855 \\
			& too few points ($\sigma$)& 0.166 & 0.172 & 0.049 & 0.044 & 0.172 & 0.085 & 0.340 & 0.364 & 0.137 \\
			& volume ($\mu$)& 1.349 & 15.641 & 50.953 & 1.291 & 14.728 & 53.616 & 2.591 & 18.698 & 58.502 \\
			& volume ($\sigma$)& 0.787 & 3.716 & 7.267 & 0.522 & 3.182 & 7.701 & 1.065 & 4.461 & 8.739 \\
			\hline	
			& realized alpha ($\mu$)& 0.900 & 0.500 & 0.100 & 0.894 & 0.510 & 0.084 & 0.903 & 0.500 & 0.095 \\
			& realized alpha ($\sigma$)& 0.000 & 0.000 & 0.000 & 0.018 & 0.038 & 0.007 & 0.008 & 0.012 & 0.012 \\
					  & too many points ($\mu$)& 0.059 & 0.082 & 0.026 & 0.043 & 0.128 & 0.030 & 0.367 & 0.280 & 0.147 \\
			$COV = A, B$ & too many points ($\sigma$)& 0.166 & 0.156 & 0.075 & 0.116 & 0.236 & 0.071 & 0.324 & 0.367 & 0.267 \\
			& too few points ($\mu$) & 0.941 & 0.893 & 0.957 & 0.952 & 0.866 & 0.954 & 0.633 & 0.674 & 0.812 \\
			& too few points ($\sigma$)& 0.166 & 0.197 & 0.109 & 0.115 & 0.235 & 0.103 & 0.324 & 0.385 & 0.306 \\
			& volume ($\mu$)& 0.432 & 5.700 & 32.295 & 0.462 & 5.882 & 35.002 & 0.828 & 7.990 & 40.587\\
			& volume ($\sigma$)& 0.196 & 1.257 & 7.993 & 0.239 & 0.978 & 9.121 & 0.339 & 2.085 & 10.329 \\
			\hline
			\end{tabular}
		\caption{Results of experiments with 10 replicates of 100 test data sets each in dimension $p=2$ for $n=100$. The data shows the average realized alpha, $\hat{\alpha}$, over all replicates and all data sets, as well as the average percentage of times generated test data has too many (or too few) points in the generated hull. Averages as labelled $\mu$ and the
                  corresponding standard devations $\sigma$.
		\label{detail_1a}}
	\end{center}
\end{table}

\begin{table} [!htbp]
	\begin{center}
		\hspace*{-2cm}  
		\begin{tabular}{| c c | r r r | r r r | r r r |}
			\hline
			& $n = 200$ && Mahal &&& Direct &&& Halfspace &  \\
			& $p=2$ & 0.9 & 0.5 & 0.1 & 0.9 & 0.5 & 0.1 & 0.9 & 0.5 & 0.1 \\
			\hline
			& realized alpha ($\mu$)& 0.900 & 0.500 & 0.100 & 0.901 & 0.503 & 0.105 & 0.901 & 0.505 & 0.106 \\
			& realized alpha ($\sigma$)& 0.000 & 0.000 & 0.000 & 0.013 & 0.022 & 0.013 & 0.002 & 0.005 & 0.010 \\
				        & too many points ($\mu$)& 0.057 & 0.063 & 0.032 & 0.081 & 0.089 & 0.025 & 0.435 & 0.363 & 0.114 \\
			$COV = A$ & too many points ($\sigma$)& 0.096 & 0.068 & 0.060 & 0.126 & 0.169 & 0.065 & 0.345 & 0.288 & 0.139 \\
			& too few points ($\mu$)& 0.943 & 0.917 & 0.954 & 0.917 & 0.898 & 0.967 & 0.565 & 0.583 & 0.851 \\
			& too few points ($\sigma$)& 0.096 & 0.080 & 0.068 & 0.125 & 0.189 & 0.084 & 0.345 & 0.289 & 0.156 \\
			& volume ($\mu$)& 0.326 & 3.678 & 10.842 & 0.339 & 2.895 & 9.162 & 0.515 & 3.365 & 10.239 \\
			& volume ($\sigma$)& 0.081 & 0.400 & 1.142 & 0.108 & 0.446 & 0.668 & 0.152 & 0.344 & 0.867 \\
			\hline
			& realized alpha ($\mu$)& 0.900 & 0.500 & 0.100 & 0.901 & 0.503 & 0.105 & 0.901 & 0.502 & 0.106 \\
			& realized alpha ($\sigma$)& 0.000 & 0.000 & 0.000 & 0.013 & 0.022 & 0.013 & 0.002 & 0.007 & 0.010 \\
				        & too many points ($\mu$)& 0.034 & 0.034 & 0.025 & 0.081 & 0.089 & 0.025 & 0.430 & 0.333 & 0.132 \\
			$COV = B$ & too many points ($\sigma$)& 0.035 & 0.034 & 0.051 & 0.126 & 0.169 & 0.065 & 0.340 & 0.296 & 0.165 \\
			& too few points ($\mu$)& 0.966 & 0.958 & 0.965 & 0.917 & 0.898 & 0.967 & 0.570 & 0.623 & 0.828 \\
			& too few points ($\sigma$)& 0.035 & 0.043 & 0.062 & 0.125 & 0.189 & 0.084 & 0.340 & 0.299 & 0.197 \\
			& volume ($\mu$)& 1.816 & 17.784 & 59.061 & 1.939 & 16.584 & 52.483 & 2.933 & 19.221 & 59.208 \\
			& volume ($\sigma$)& 0.377 & 1.494 & 6.116 & 0.620 & 2.558 & 3.825 & 0.855 & 1.961 & 4.711 \\
			\hline	
			& realized alpha ($\mu$)& 0.900 & 0.500 & 0.100 & 0.896 & 0.494 & 0.100 & 0.902 & 0.502 & 0.103 \\
			& realized alpha ($\sigma$)& 0.000 & 0.000 & 0.000 & 0.015 & 0.018 & 0.008 & 0.003 & 0.006 & 0.009 \\
					  & too many points ($\mu$)& 0.080 & 0.034 & 0.050 & 0.096 & 0.066 & 0.021 & 0.343 & 0.370 & 0.149 \\
			$COV = A, B$ & too many points ($\sigma$)& 0.155 & 0.046 & 0.118 & 0.200 & 0.098 & 0.043 & 0.372 & 0.281 & 0.249 \\
			& too few points ($\mu$)& 0.920 & 0.949 & 0.937 & 0.904 & 0.914 & 0.972 & 0.657 & 0.593 & 0.811 \\
			& too few points ($\sigma$)& 0.155 & 0.069 & 0.138 & 0.200 & 0.118 & 0.061 & 0.372 & 0.290 & 0.257 \\
			& volume ($\mu$)& 0.557 & 6.173 & 37.733 & 0.584 & 6.879 & 37.073 & 0.831 & 8.152 & 42.935 \\
			& volume ($\sigma$)& 0.172 & 0.751 & 4.970 & 0.168 & 0.896 & 3.130 & 0.263 & 0.843 & 5.534 \\
			\hline	
		\end{tabular}
		\caption{Results of experiments with ten replicates of 100 test data sets each in dimension $p=2$ for $n=200$. The data shows the average realized alpha over all replicates and all data sets, as well as the average percentage of times generated test data has too many (or too few) points in the generated hull. Averages as labelled $\mu$ and the
                  corresponding standard devations $\sigma$.
		\label{detail_1b}}
	\end{center}
\end{table}

\begin{table} [!htbp]
	\begin{center}
		\hspace*{-2.5cm}  
		\begin{tabular}{| c c | r r r | r r r | r r r |}
			\hline
			& $n = 500$ && Mahal &&& Direct &&& Halfspace &  \\
			& $p=2$ & 0.9 & 0.5 & 0.1 & 0.9 & 0.5 & 0.1 & 0.9 & 0.5 & 0.1 \\
			\hline
			& realized alpha ($\mu$)& 0.900 & 0.500 & 0.100 & 0.894 & 0.501 & 0.104 & 0.901 & 0.501 & 0.102 \\
			& realized alpha ($\sigma$)& 0.000 & 0.000 & 0.000 & 0.004 & 0.013 & 0.012 & 0.001 & 0.002 & 0.003 \\
				        & too many points ($\mu$)& 0.164 & 0.114 & 0.086 & 0.120 & 0.090 & 0.027 & 0.511 & 0.307 & 0.285 \\
			$COV = A$ & too many points ($\sigma$)& 0.302 & 0.090 & 0.129 & 0.183 & 0.151 & 0.028 & 0.321 & 0.297 & 0.245 \\
			& too few points ($\mu$)& 0.836 & 0.859 & 0.902 & 0.880 & 0.898 & 0.965 & 0.489 & 0.667 & 0.675 \\
			& too few points ($\sigma$)& 0.302 & 0.098 & 0.143 & 0.183 & 0.169 & 0.034 & 0.321 & 0.299 & 0.260 \\
			& volume ($\mu$)& 0.443 & 4.242 & 12.669 & 0.453 & 3.122 & 10.148 & 0.520 & 3.373 & 11.178 \\
			& volume ($\sigma$)& 0.115 & 0.225 & 0.917 & 0.072 & 0.192 & 0.639 & 0.084 & 0.220 & 0.503 \\
			\hline
			& realized alpha ($\mu$)& 0.900 & 0.500 & 0.100 & 0.894 & 0.501 & 0.104 & 0.901 & 0.500 & 0.102 \\
			& realized alpha ($\sigma$)& 0.000 & 0.000 & 0.000 & 0.004 & 0.013 & 0.012 & 0.001 & 0.003 & 0.004 \\
				        & too many points ($\mu$)& 0.204 & 0.102 & 0.045 & 0.120 & 0.090 & 0.027 & 0.495 & 0.312 & 0.262 \\
			$COV = B$ & too many points ($\sigma$)& 0.348 & 0.172 & 0.070 & 0.183 & 0.151 & 0.028 & 0.315 & 0.282 & 0.242 \\
			& too few points ($\mu$)& 0.796 & 0.890 & 0.938 & 0.880 & 0.898 & 0.965 & 0.505 & 0.662 & 0.699 \\
			& too few points ($\sigma$)& 0.348 & 0.176 & 0.096 & 0.183 & 0.169 & 0.034 & 0.315 & 0.291 & 0.266 \\
			& volume ($\mu$)& 2.545 & 20.121 & 68.944 & 2.592 & 17.882 & 58.129 & 2.978 & 19.350 & 63.699 \\
			& volume ($\sigma$)& 0.732 & 1.430 & 5.305 & 0.412 & 1.098 & 3.662 & 0.484 & 1.267 & 3.212 \\
			\hline	
			& realized alpha ($\mu$)& 0.900 & 0.500 & 0.100 & 0.904 & 0.501 & 0.102 & 0.900 & 0.501 & 0.101 \\
			& realized alpha ($\sigma$)& 0.000 & 0.000 & 0.000 & 0.007 & 0.009 & 0.009 & 0.002 & 0.002 & 0.004 \\
				             & too many points ($\mu$)& 0.148 & 0.087 & 0.047 & 0.220 & 0.057 & 0.032 & 0.545 & 0.284 & 0.301 \\
			$COV = A, B$ & too many points ($\sigma$)& 0.173 & 0.171 & 0.058 & 0.304 & 0.103 & 0.049 & 0.340 & 0.297 & 0.241 \\
			& too few points ($\mu$)& 0.852 & 0.900 & 0.936 & 0.780 & 0.928 & 0.961 & 0.455 & 0.686 & 0.656 \\
			& too few points ($\sigma$)& 0.173 & 0.197 & 0.079 & 0.304 & 0.120 & 0.059 & 0.340 & 0.306 & 0.262 \\
			& volume ($\mu$)& 0.762 & 6.826 & 43.891 & 0.756 & 7.077 & 41.022 & 0.985 & 7.996 & 46.339 \\
			& volume ($\sigma$)& 0.125 & 0.683 & 3.497 & 0.154 & 0.664 & 3.874 & 0.168 & 0.670 & 3.057 \\
			\hline	
		\end{tabular}
		\caption{Results of experiments with ten replicates, dimension $p=2$ for $n=500$. The data shows the average realized alpha over all replicates and all data sets, and the average percentage of times generated test data has too many (or too few) points in the generated hull. Averages as labelled $\mu$ and the
                  corresponding standard devations $\sigma$.
		\label{detail_2a}}
	\end{center}
\end{table}

\begin{table} [!htbp]
	\begin{center}
		\hspace*{-2.5cm}  
		\begin{tabular}{| c c | r r r | r r r | r r r |}
			\hline	
			& $n = 200$ && Mahal &&& Direct &&& Halfspace &  \\
			& $p=3$ & 0.9 & 0.5 & 0.1 & 0.9 & 0.5 & 0.1 & 0.9 & 0.5 & 0.1 \\
			\hline
			& realized alpha ($\mu$)& 0.900 & 0.500 & 0.100 & 0.887 & 0.484 & 0.128 & 0.903 & 0.522 & 0.029 \\
			& realized alpha ($\sigma$)& 0.000 & 0.000 & 0.000 & 0.029 & 0.045 & 0.015 & 0.011 & 0.038 & 0.062 \\
				        & too many points ($\mu$)& 0.000 & 0.002 & 0.000 & 0.000 & 0.000 & 0.000 & 0.665 & 0.920 & 0.010 \\
			$COV = A$ & too many points ($\sigma$)& 0.000 & 0.004 & 0.000 & 0.000 & 0.000 & 0.000 & 0.386 & 0.116 & 0.032 \\
			& too few points ($\mu$)& 1.000 & 0.996 & 1.000 & 1.000 & 1.000 & 1.000 & 0.335 & 0.061 & 0.986 \\
			& too few points ($\sigma$)& 0.000 & 0.007 & 0.000 & 0.000 & 0.000 & 0.000 & 0.386 & 0.096 & 0.041 \\
			& volume ($\mu$)& 0.354 & 8.042 & 23.520 & 0.421 & 5.303 & 18.115 & 1.375 & 11.950 & 33.960 \\
			& volume ($\sigma$)& 0.064 & 1.324 & 2.427 & 0.211 & 1.107 & 2.241 & 0.438 & 1.728 & 5.974 \\
			\hline
			& realized alpha ($\mu$)& 0.900 & 0.500 & 0.100 & 0.887 & 0.484 & 0.128 & 0.897 & 0.515 & 0.081 \\
			& realized alpha ($\sigma$)& 0.000 & 0.000 & 0.000 & 0.029 & 0.045 & 0.015 & 0.010 & 0.020 & 0.086 \\
				        & too many points ($\mu$)& 0.000 & 0.000 & 0.000 & 0.000 & 0.000 & 0.000 & 0.645 & 0.664 & 0.055 \\
			$COV = B$ & too many points ($\sigma$)& 0.000 & 0.000 & 0.000 & 0.000 & 0.000 & 0.000 & 0.292 & 0.352 & 0.160 \\
			& too few points ($\mu$)& 1.000 & 0.999 & 1.000 & 1.000 & 1.000 & 1.000 & 0.351 & 0.308 & 0.929 \\
			& too few points ($\sigma$)& 0.000 & 0.003 & 0.000 & 0.000 & 0.000 & 0.000 & 0.295 & 0.345 & 0.190 \\
			& volume ($\mu$)& 4.417 & 75.264 & 251.685 & 4.966 & 62.567 & 213.727 & 15.693 & 117.871 & 358.733 \\
			& volume ($\sigma$)& 1.711 & 12.566 & 22.361 & 2.493 & 13.066 & 26.442 & 3.851 & 15.147 & 65.221 \\
			\hline	
			& realized alpha ($\mu$)& 0.900 & 0.500 & 0.100 & 0.901 & 0.505 & 0.110 & 0.899 & 0.516 & 0.116 \\
			& realized alpha ($\sigma$)& 0.000 & 0.000 & 0.000 & 0.027 & 0.030 & 0.016 & 0.009 & 0.060 & 0.065 \\
				             & too many points ($\mu$)& 0.002 & 0.000 & 0.000 & 0.000 & 0.000 & 0.000 & 0.600 & 0.769 & 0.096 \\
			$COV = A, B$ & too many points ($\sigma$)& 0.004 & 0.000 & 0.000 & 0.000 & 0.000 & 0.000 & 0.333 & 0.307 & 0.096 \\
			& too few points ($\mu$)& 0.998 & 1.000 & 1.000 & 1.000 & 1.000 & 1.000 & 0.400 & 0.213 & 0.870 \\
			& too few points ($\sigma$)& 0.004 & 0.000 & 0.000 & 0.000 & 0.000 & 0.000 & 0.333 & 0.296 & 0.118 \\
			& volume ($\mu$)& 0.686 & 15.941 & 174.555 & 0.735 & 20.455 & 173.824 & 2.634 & 71.708 & 275.968 \\
			& volume ($\sigma$)& 0.232 & 2.324 & 22.007 & 0.518 & 5.692 & 32.283 & 0.711 & 50.054 & 72.006  \\
			\hline	
		\end{tabular}
		\caption{Results of experiments with ten replicates in dimension $p=3$ for $n=200$. The data shows the average realized alpha out of all replicates and all data sets, and the average percentage of times generated test data has too many (or too few) points in the generated hull. Averages as labelled $\mu$ and the
                  corresponding standard devations $\sigma$.
		\label{detail_2b}}
	\end{center}
\end{table}

\begin{table} [ht]
	\centering
	\hspace*{-1cm} 
	\begin{tabular}{| c | c | c | r r r | r r r | r r r |}
		\hline
		runtime && && Mahal &&& Direct &&& Halfspace &  \\
		$p$ & $n$ & $COV$ & 0.9 & 0.5 & 0.1 & 0.9 & 0.5 & 0.1 & 0.9 & 0.5 & 0.1 \\
		\hline
		&& $A$ & 14.9 & 14.6 & 15.1 & 15.1 & 14.8 & 15.6 & 50.5 & 42.1 & 39.8 \\
		&$n=100$& $B$ & 19.4 & 15.2 & 15.4 & 15.2 & 15.9 & 15.2 & 51.9 & 43.1 & 39.5 \\
		&& $A, B$ & 15.7 & 14.9 & 15.0 & 15.3 & 15.0 & 15.5 & 51.3 & 42.8 & 39.0 \\
		\cline{2-12}
		&& $A$ & 16.6 & 17.9 & 17.3 & 16.8 & 16.7 & 16.9 & 277.8 & 252.7 & 244.4 \\
		$p=2$& $n=200$ & $B$ & 17.3 & 16.8 & 16.8 & 17.1 & 17.1 & 16.5 & 297.3 & 255.4 & 239.4 \\
		&& $A, B$ & 17.2 & 16.5 & 18.0 & 16.7 & 16.6 & 17.3 & 315.9 & 242.8 & 234.3 \\
		\cline{2-12}
		&& $A$ & 16.9 & 16.9 & 17.2 & 17.4 & 16.6 & 17.1 & 3585.1 & 3228.5 & 3110.8 \\
		& $n=500$ & $B$ & 16.8 & 17.4 & 17.5 & 17.0 & 17.1 & 17.1 & 3538.4 & 3255.9 & 3065.9 \\
		&& $A, B$ & 17.2 & 17.2 & 20.7 & 17.2 & 16.8 & 17.0 & 3509.1 & 3318.3 & 3094.6 \\
		\hline
		&& $A$ & 24.0 & 25.3 & 23.4 & 23.2 & 22.6 & 23.2 & 18668.8 & 17187.6 & 18271.0 \\
		$p=3$&$n=200$& $B$ & 23.6 & 22.8 & 23.2 & 22.8 & 22.6 & 22.5 & 18233.0 & 17267.5 & 18010.7 \\
		&& $A, B$ & 25.1 & 26.2 & 24.5 & 22.6 & 23.8 & 23.4 & 18171.9 & 17233.8 & 17076.9 \\
		\hline
	\end{tabular}
	\caption{Total runtime in seconds of the construction time of ten replicates for the experiments summarized in \autoref{big_table} and \autoref{3d_table}. These experiments were run on a 64-Bit UNIX system with an \texttt{$\mbox{Intel}_{\mbox{\textregistered}}$ $\mbox{Core}^{\mbox{\texttrademark}}$ i7 CPU 860 @ 2.80GHz} processor.}
	\label{time_table}
\end{table}

\end{document}